\documentclass[11pt]{article}
\usepackage{amssymb, latexsym, mathtools, amsthm}
\begingroup
    \makeatletter
    \@for\theoremstyle:=definition,remark,plain\do{%
        \expandafter\g@addto@macro\csname th@\theoremstyle\endcsname{%
            \addtolength\thm@preskip\parskip
            }%
        }
\endgroup
\usepackage{enumerate}
\usepackage{pgf,tikz}
\usepackage{pdfsync}
\usepackage{txfonts}
\usepackage[T1]{fontenc}
\usepackage{booktabs}
 \usepackage{graphicx}
\definecolor{dnrbl}{rgb}{0,0,0.3}
\definecolor{dnrgr}{rgb}{0,0.3,0}
\definecolor{dnrre}{rgb}{0.5,0,0}
\usepackage[colorlinks=true, citecolor=dnrgr, linkcolor=dnrre, urlcolor=dnrre] {hyperref}
\usepackage{xcolor}
\usepackage{parskip}
\usepackage{tabularx, colortbl}
\usepackage{geometry}
\usepackage{etoolbox}
 \usepackage[scriptsize, up]{caption}
\usepackage{pgf,tikz}
\usetikzlibrary{decorations, decorations.pathmorphing}
\usetikzlibrary {shapes}
\theoremstyle{plain}
\newtheorem{thm}{Theorem}[section]

\newtheorem{lem}[thm]{Lemma}
\newtheorem{coro}[thm]{Corollary}

\newtheorem{defi}[thm]{Definition}
\makeatletter

\let\c@table\c@figure
\makeatother

\newcommand{\Nat}{\mathbb{N}}

\newcommand{\restr}{\upharpoonright}  

\newcommand{\smo}[1]{\mathop{\bf o}\/\left({#1}\right)}
\DeclarePairedDelimiter{\dbra}{\llbracket}{\rrbracket}
\DeclarePairedDelimiter{\dvls}{\lVert}{\rVert}

\newcommand{\DD}{\mathcal{D}}

\newcommand{\wgt}[1]{\mathop{\mathtt{wgt}}\/\left({#1}\right)}

\newcommand{\TTast}{\mathcal{T}^{\ast}} 
\newcommand{\TT}{\mathcal{T}}

\newcommand{\PP}{\mathcal{P}} 
\newcommand{\QQ}{\mathcal{Q}} 
 
\newcommand{\MM}{Merkle and Mihailovi\'{c}\ }

\newcommand{\ml}{Martin-L\"{o}f }
\newcommand{\KG}{Ku\v{c}era-G{\'a}cs\ }
\newcommand{\pz}{$\Pi^0_1$\ }

\newcommand{\ie}{i.e.\ }
\newcommand{\ce}{c.e.\ }

\renewenvironment{abstract}
 { \normalsize
  \list{}{
    \setlength{\leftmargin}{.0cm}%
    \setlength{\rightmargin}{\leftmargin}%
    }%
  \item {\bf \abstractname.} \relax}
 {\endlist}
\usepackage{titlesec}
 \titleformat{\paragraph}
{\normalfont\normalsize\em\bfseries}{\theparagraph}{1em}{$\blacktriangleright$\hspace{0.2cm}}
\titlespacing*{\paragraph}
{0pt}{3.25ex plus 1ex minus .2ex}{0.5ex plus .2ex}

\title{Optimal redundancy in computations from random oracles
\thanks{Barmpalias was supported by the 
1000 Talents Program for Young Scholars from the Chinese Government, grant no.\ D1101130.
Additional support was received by
the Chinese Academy of Sciences (CAS) and the Institute of Software of the CAS.
Lewis-Pye was supported by a Royal Society University 
Research Fellowship.}}

\author{George Barmpalias  \and Andrew Lewis-Pye}
\date{\today}
\begin{document}
\maketitle
\begin{abstract}
It is a classic result in algorithmic information theory that
every infinite binary sequence is computable from an infinite binary sequence which is random in the sense of Martin-L\"{o}f. 
Proved independently by
Ku\v{c}era \cite{MR820784} and G{\'a}cs \cite{MR859105}, this result answered a question by Charles Bennett and
has seen numerous applications in the last 30 years. 
The optimal redundancy in such a coding process has, however,  remained unknown.
If the computation of the first $n$ bits of a sequence requires $n+g(n)$ bits 
of the random oracle, then $g$ is the {\em redundancy} of the computation.
Ku\v{c}era implicitly achieved redundancy $n\log n$ while G{\'a}cs  used a more elaborate block-coding procedure
which achieved redundancy $\sqrt{n}\log n$. 
Merkle and Mihailovi\'{c} \cite{jsyml/MerkleM04} provided a different presentation of
G{\'a}cs' approach, without improving his redundancy bound.
In this paper we devise a new coding method that achieves optimal logarithmic redundancy.
For any computable non-decreasing function $g$ such that $\sum_i 2^{-g(i)}$ is bounded we show that
there is a coding process that codes any given 
infinite binary sequence into a \ml random infinite binary sequence with redundancy $g$.
This redundancy bound is exponentially smaller than the previous bound of
$\sqrt{n}\log n$ and
is known to be the best possible by recent work \cite{lbrecorao15}, where it was shown that if  $\sum_i 2^{-g(i)}$ diverges then there exists an infinite binary sequence $X$ which cannot be computed by any \ml random infinite binary sequence with redundancy $g$. 
It follows that redundancy $\epsilon\cdot\log n$ in
computation from a random oracle is possible
for every infinite binary sequence, if and only if $\epsilon>1$. 
\end{abstract}
\vspace*{\fill}
\noindent{\bf George Barmpalias}\\[0.5em]
\noindent
State Key Lab of Computer Science, 
Institute of Software, Chinese Academy of Sciences, Beijing, China.\\[0.2em] 
\textit{E-mail:} \texttt{\textcolor{dnrgr}{barmpalias@gmail.com}}.
\textit{Web:} \texttt{\textcolor{dnrre}{http://barmpalias.net}}\par
\addvspace{\medskipamount}\medskip\medskip
\noindent{\bf Andrew Lewis-Pye}\\[0.5em]  
\noindent Department of Mathematics,
Columbia House, London School of Economics, 
Houghton Street, London, WC2A 2AE, United Kingdom.\\[0.2em]
\textit{E-mail:} \texttt{\textcolor{dnrgr}{A.Lewis7@lse.ac.uk.}}
\textit{Web:} \texttt{\textcolor{dnrre}{http://aemlewis.co.uk}} 

\vfill \thispagestyle{empty}
\clearpage

\section{Introduction}
If an infinite binary sequence is algorithmically random, one does not expect to be able to extract any useful information from it.
Although this reasonable intuition can be verified in many formal contexts\footnote{If an 
infinite binary sequence is
{\em arithmetically random} in the sense that 
it avoids all null sets of reals which are arithmetically definable, then it cannot compute any
noncomputable infinite binary sequence that is definable in arithmetic. Similarly, if an infinite binary sequence is \ml random relative to
the halting problem, then it cannot compute any noncomputable infinite binary sequence which is 
definable in arithmetic with one
unbounded quantifier. A less trivial example is a fact from Stephan 
\cite{MR2258713frank} that incomplete \ml random
infinite binary sequences  
cannot compute any complete extensions of Peano Arithmetic, and its extensions in 
Levin \cite{focs/Levin02,Levin:2013:FI}.}, it fails for the most accepted and robust 
notion of algorithmic randomness,  which is \ml randomness \cite{MR0223179},
also formulated as incompressibility in terms of Kolmogorov complexity by Chaitin \cite{MR0411829}
and Levin \cite{MR0366096}.
Indeed, Ku\v{c}era \cite{MR820784}, and independently G{\'a}cs \cite{MR859105}, showed
that any infinite binary sequence is computable from a \ml random sequence.
Both authors constructed a uniform process that codes every infinite binary sequence
into some \ml random infinite binary sequence.
The \KG theorem, as it is known in algorithmic information theory, has been studied and extended in numerous ways
in the last 30 years\footnote{For example see 
\cite{Kucera:87,bookrandsiam,jucs/Hertling97,jsyml/MerkleM04,cie/Doty06,MR2170569,BDNGP}.}
and has become a standard prominent topic in most textbooks and presentations of this area.\footnote{For example
consider the standard textbooks \cite{Li.Vitanyi:93,Calude:94,Ottobook,rodenisbook} and
the surveys \cite{MR2248591,MR2248590}.} 
In the context of \ml randomness, this result says that:
\[
\parbox{14cm}{{\small Any type of information that can be coded into an infinite binary sequence, 
no matter how structured that might be, can
be obfuscated into an algorithmically random infinite binary sequence, 
from which it is effectively recoverable.}}
\]
Here {\em information} could be the solution to a problem of interest, such as the halting problem, the word problem for
finite groups, or any of the numerous and often algorithmically unsolvable problems whose solutions can be represented
as  a set of integers. {\em Effectively recoverable} means computable by means of a Turing reduction, without any
restrictions on time or memory. As we discuss below, however, the coding constructed in both
 \cite{MR820784} and \cite{MR859105} gives a Turing reduction with a computable upper bound on the length
 of the initial segment of the oracle that is used in the computation on any given argument---the {\em oracle use}.\footnote{In the
 terminology of computability theory, every infinite binary sequence is weak-truth-table computable from a \ml random
infinite binary sequence. Bennett \cite{Bennett1988} observed that this is no longer true for truth-table computations, and used this
 fact in order to define {\em logical depth} for infinite binary sequences. Other refined reducibilities were considered
 by Book \cite{bookrandsiam}.}

It is hardly surprising that such a coding process occasionally introduces an overhead on the codes of the initial
segments of certain infinite binary sequences. More specifically, if we code an infinite binary sequence $X$ into a \ml random infinite binary sequence $Y$, then it is
very possible that for some $n$, in order to recover the first $n$ bits of $X$ (denoted $X\restr_n$), we need $Y\restr_{n+g(n)}$,
\ie $g(n)$ more bits of $Y$. Such a function $g$ that bounds from above the number of extra bits needed in the decoding process
is known as the {\em redundancy} in the computation of $X$ from $Y$.
For example, it is known from \cite{lbrecorao15} that certain infinite binary sequences $X$ are not computable from any
\ml random infinite binary sequence with redundancy $\log n$. Such restrictions can be intuitively understood if one considers that information
introduces structure, and in order to obfuscate the structure of a given $X$ into a random $Y$, extra bits amplifying the
complexity of the code might be necessary.

In the context of information theory, it is important to: 
\begin{enumerate}[\hspace{0.5cm}(a)]
\item Determine the optimal redundancy in coding into \ml random infinite binary sequences;
\item Construct a coding process that achieves the optimal redundancy.
\end{enumerate}
The original work in \cite{MR820784,MR859105} did not achieve these goals, nor did the subsequent
work of Merkle and Mihailovi\'{c} \cite{jsyml/MerkleM04} and Doty \cite{cie/Doty06}. 
The goal of the present work is to give a definitive answer to challenges (a) and (b).

\subsection{Previous, directly relevant work}
Ku\v{c}era \cite{MR820784} did not show an interest in optimising  the redundancy of his coding, other than observing
that it can be computably bounded. An examination of his argument (see the survey \cite{codico16} for such a discussion)
shows that his method produces redundancy $n\log n$.
G\'{a}cs \cite{MR859105}, on the other hand, has a clear interest in minimising the redundancy of his coding, which
he bounds by $\sqrt{n}\log n$ by means of a more sophisticated block-coding, with carefully chosen 
block-lengths.\footnote{In actual fact, G\'{a}cs \cite{MR859105} achieves redundancy $3\sqrt{n}\log n$, but a careful examination of his argument (as this is discussed in the survey  \cite{codico16}) shows that it can be reduced to 
$\sqrt{n}\log n$. }
\MM \cite{jsyml/MerkleM04} give an interpretation of  G\'{a}cs' coding in terms of effective martingales, instead
of the effective closed sets approach employed in the original argument. Although the latter analysis is rather
elegant and geared
toward obtaining a small redundancy $\smo{n}$,
the resulting upper bound is identical with G\'{a}cs' bound of$\sqrt{n}\log n$.

Doty \cite{cie/Doty06} showed how to reduce the oracle-use when coding an infinite binary sequence $X$ into a \ml random $Y$,
based on suitable bounds on the constructive dimension of $X$. Partially extending
previous work by Ryabko \cite{Rya86}, he showed that the
asymptotic ratio between the optimal 
oracle-use in computing $X\restr_n$ and $n$ is directly related to the constructive dimension of $X$.
Unfortunately, the arguments developed in this latter work do not shed light on our main goal, which asks 
for the {\em actual} optimal redundancy in coding into \ml random infinite binary sequences, and not the mere asymptotic behavior of
the oracle-use in such a reduction. For infinite binary sequences $X$ of dimension 1, for example, the work in \cite{cie/Doty06}
merely shows that they can be computed by a \ml random infinite binary sequence $Y$ with oracle-use $\ell_n$ such that 
$\liminf_n (\ell_n/n)=1$. From the latter we cannot even deduce G\'{a}cs upper 
bound $n+\sqrt{n}\log n$ on the oracle-use.

\subsection{Our results}
Our main contribution is a coding process that codes an arbitrary infinite binary sequence into a \ml random infinite binary sequence,
with optimal redundancy which is exponentially smaller\footnote{When $f$ and $g$ are unbounded, we say $g$ is exponentially smaller than $f$ if there exists a constant $c$ such that $2^{c\cdot g(n)}<f(n)$ for all $n$.} than the previous known bound of $\sqrt{n}\log n$.
In the following statement, `uniformly computable' means that there is a single coding process that works for all infinite binary sequences, \ie a single Turing functional that provides the promised reduction  of 
each given infinite binary sequence to some \ml random infinite binary sequence.

\begin{thm}\label{Da7UFTxKv}
If $(\ell_i)$ is a  computable increasing function such that $\sum_i 2^{-\ell_i+i}<1$ then every  
infinite binary sequence is
uniformly computable from a \ml random infinite binary sequence with oracle-use $(\ell_i)$.
\end{thm}
If we require only that $\sum_i 2^{-\ell_i+i}$ is bounded in the statement of Theorem \ref{Da7UFTxKv}, then of course the same conclusion will hold so long as either: (a) we allow oracle use $(\ell_i)+c$ for some constant $c$, or (b) we drop the requirement of uniformity.  In fact, Theorem \ref{Da7UFTxKv} is part of the following 
slightly more general fact that we prove, regarding coding into
effectively closed sets of positive measure.
\begin{lem}\label{Dedk3zkhlL}
Let $(\ell_i)$ be an increasing computable sequence and let $\PP$ be a \pz class.
If $\sum_i 2^{-\ell_i+i}<\mu(\PP)$ then every infinite binary sequence is uniformly computable from
some member of $\PP$ with oracle-use $(\ell_i)$.
\end{lem}
Note that Theorem  \ref{Da7UFTxKv} follows directly from Lemma \ref{Dedk3zkhlL}
since the class of \ml random infinite binary sequences is a $\Sigma^0_2$ set of measure 1.
We are also able to establish the optimality of these two results.
In \cite{lbrecorao15} it was shown that if the sum in Theorem \ref{Da7UFTxKv} is not bounded, 
then there exists an infinite binary sequence
which is not computable by any \ml random infinite binary sequence with oracle-use $(\ell_i)$.
If we combine this with Theorem  \ref{Da7UFTxKv} we get the following characterization.
\begin{coro}\label{YtJV9pxCc7}
Let $g$ be a nondecreasing computable function. Then  the following are equivalent:
\begin{enumerate}[\hspace{0.5cm}(i)]
\item every infinite binary sequence  is computable from a \ml random infinite binary sequence with redundancy $g$;
\item  $\sum_i 2^{-g(i)}<\infty$.
\end{enumerate}
\end{coro}
Note that in clause (i) for Corollary \ref{YtJV9pxCc7} we can replace `computable' with `uniformly computable' so long as the sum in (ii) is strictly bounded by 1.

\subsection{Terminology, methodology and novelty}

\subsubsection{Terminology}\label{VLeuwFyfem}
The Cantor space $2^{\omega}$ is the class of all infinite binary sequences. Unless explicitly stated otherwise, it is to be assumed that binary strings are finite, so that strings are finite objects. We let $|\sigma|$ denote the length of the string $\sigma$. If $Q$ is a computably enumerable
set of binary strings, we let $\dbra{Q}$ denote the class of infinite binary sequences which are prefixed by some string in $Q$.
If $Q=\{\nu\}$, then we simply write $\dbra{\nu}$ for $\dbra{Q}$.
In this way $\Sigma^0_1$ subsets of  $2^{\omega}$ can be represented by \ce sets of strings $Q$.
The Lebesgue measure of $\dbra{Q}$ may be denoted simply by $\mu(Q)$.
A tree $T$ in the Cantor space is a downward closed set of binary strings with respect to the prefix relation.
A branch of $T$ is simply a string in $T$ and a path through $T$ is an infinite binary sequence for which all finite initial segments
are branches of $T$. 
A \pz class in the Cantor space can be represented as a $\Pi^0_1$ tree or as 
$2^{\omega}-\dbra{Q}$ for some \ce set of strings $Q$.
 The {\em $n$-th level} of $T$ consists of the strings in $T$ of length $n$.
 A {\em leaf} of a tree is a branch of the tree with no proper extensions in the tree. 
For any set of strings $T$,  we let $[T]$ denote the set of all infinite binary sequences with infinitely many
prefixes in $T$. Note that when $T$ is a tree, $[T]$ denotes the set of infinite paths through $T$.

Now suppose we are given the \pz class $\PP$ and the increasing computable
sequence $(\ell_i)$ of Lemma \ref{Dedk3zkhlL}. Our task is to construct a Turing functional $\Phi$ with uniform oracle-use $(\ell_i)$ 
on all oracles, with the property that for every infinite binary sequence $X$ there exists some $Y_X\in\PP$ such
that $X=\Phi^{Y_X}$.
It is convenient to define $\Phi$ by assigning labels for strings of each length $n$ to strings of each length $\ell_n$. If $\sigma$ is of length $n$, $\tau$ is of length $\ell_n$ and we assign the label $x_{\sigma}$ to $\tau$, then this is equivalent to defining $\Phi^{\tau}=\sigma$.    Of course these assignments
need to be consistent, in the sense that if 
$\tau\subseteq\tau'$,  $\Phi^{\tau}=\sigma$ and  $\Phi^{\tau'}=\sigma'$ then
$\sigma\subseteq\sigma'$. In our analysis we thus present $\Phi$ as a \emph{partially labelled tree}, by which we mean the full binary tree $2^{<\omega}$ along with a partial labelling of it.
Given a partially labelled tree $\TT$ and $\ell\in\Nat$, we let 
 $\TT\restr_{\ell}$ denote the restriction of $\TT$ to the strings of length at most $\ell$.

\subsubsection{The departure from existing approaches}
 There are a number
of different presentations of the \KG theorem in the literature. Ku\v{c}era \cite{MR820784} uses the
recursion theorem and the universality properties of the class of \ml random infinite binary sequences. His coding
method may be seen as being of the following inductive form.  Working within a $\Pi^0_1$ class of Martin-L\"{o}f randoms $\mathcal{P}$, which is the set of all infinite paths through the computable tree $\mathcal{T}$, let us suppose that we have already determined $2^n$ strings of length $\ell_n$ in $\mathcal{T}$ which are extendable (i.e. have infinite extensions in $\mathcal{P}$), such that for each string $\sigma$ of length $n$ there is precisely one of these extendable strings $\tau$ for which we have defined $\Phi^{\tau}=\sigma$. From properties of the class $\mathcal{P}$, we are then able to determine a length $\ell_{n+1}$ such that each of these $2^n$ strings $\tau$ must have at least two incompatible and extendable extensions in $\mathcal{T}$ of length $\ell_{n+1}$. If $\Phi^{\tau}=\sigma$, then for two of these extendable extensions $\tau'$ and $\tau''$ of length $\ell_{n+1}$,  we can define $\Phi^{\tau'}=\sigma \ast 0$ and $\Phi^{\tau''}=\sigma \ast 1$. The coding may therefore be thought of as occurring \emph{bit-by-bit},  and actually takes place inside  a subclass $\PP'\subseteq\PP$ defined by the tree $\mathcal{T}'$ with the property that for all $n$: 
\begin{equation}\label{FtwiAIkrM4}
\parbox{13cm}{Every branch of $\mathcal{T}'$ at level $\ell_n$ has at least two extensions at
level $\ell_{n+1}$ in $\mathcal{T}'$.}
\end{equation}
As we proceed to code $X$, the manner in which we code $\sigma \ast i\subset X$ may also be seen to satisfy a strong \emph{independence} property:  our code for the initial segment of $X$ which is $\sigma \ast i$  depends only on $\mathcal{P}$, $i$,  and the code for $\sigma$ (and not, for example, on $X(n)$ for $n>|\sigma \ast i |$).  

In G\'{a}cs' approach, he does not code bit-by-bit, but rather breaks the infinite binary sequences to be coded into finite blocks of appropriately chosen lengths, and then codes each block rather than each bit one at a time. 
Coding in blocks in this way allows for a substantial reduction in the redundancy. Nevertheless, it is easily seen that weaker versions of the independence property and condition (\ref{FtwiAIkrM4}) still hold. If the $(n+1)$st block is of length $m_n$ then (\ref{FtwiAIkrM4}) will hold with two replaced by $2^{m_n}$. Similarly, the way in which we code the $(n+1)$st block will depend only on $\mathcal{P}$ and the coding of previous blocks.   
In order to achieve an exponentially smaller redundancy bound with our coding, we shall need to develop more general techniques, for which neither of these strong restrictions apply. 

\subsection{Background and organization}
We assume a basic working knowledge of computability theory and its main concepts.
Other than that, the proof of Lemma \ref{Dedk3zkhlL} is self-contained. In particular, knowledge
of previous proofs of the \KG theorem is not assumed. The reader who is interested in 
a more detailed analysis of the different approaches to the task of coding into random infinite binary sequences,
is referred to the recent survey \cite{codico16}.
 For background on \ml randomness we refer to the textbooks
Li and Vitanyi \cite{Li.Vitanyi:93}, Downey and Hirschfeldt \cite{rodenisbook} or 
Nies \cite{Ottobook}. The latter two books also contain background in 
computability theory.

As we discussed in Section \ref{VLeuwFyfem}, the promised reduction of Lemma \ref{Dedk3zkhlL}
will be achieved by means of a labelling of the full binary tree. Section
\ref{mtzrcubXzt} is devoted to the construction of this labelling and the 
statement of its key properties. In Section \ref{EgBPR7blWw} and Section \ref{yPyARDIifv} we verify the properties
of the labelling construction  and complete the proof of 
 Lemma  \ref{Dedk3zkhlL}.  

\section{Partial labelling of the full binary tree}\label{mtzrcubXzt}
The reduction needed for the proof of Lemma  \ref{Dedk3zkhlL} is
constructed via the enumeration of a partially labelled tree $\TT$ with certain properties, which we construct in this section.
Recall that we are given a \pz class $\PP$ and an increasing  computable sequence $(\ell_i)$ such that:
\begin{equation}\label{bi1VeHzCL9}
 \sum_i 2^{-\ell_i+i}<\mu(\PP). 
\end{equation}
The partially labelled tree $\TT$ will be determined as the limit of a computable sequence
$(\TT_s)$ of partially labelled trees. We call
$(\TT_s)$ a {\em labelling process} for $\TT$. Let $\QQ$ be a \ce set of binary strings such that 
$\PP=2^{\omega}-\dbra{Q}$. Let $(\QQ_s)$ be a computable enumeration of $\QQ$.
Before we give the construction of $(\TT_s)$, we state a number of key properties that  $(\TT_s)$ will have
and define some relevant notions. 

\subsection{Basic properties of the labelling}\label{RBAKF1pPzY}
The partially labelled tree $\TT$ that we construct will be structured in the following sense.

\begin{defi}[Structured partially labelled trees]\label{oFUyKrVWUK}
A partially labelled tree $\TT$ is  structured with respect to an increasing sequence $(\ell_i)$,  
if the following properties are met.
\begin{enumerate}[\hspace{0.5cm}(1)]
\item {\em Restriction:} only strings at levels $\ell_i, i\in\Nat$ of $\TT$ can have a label;
\item {\em Layering:} the labels placed on the level $\ell_i$ of $\TT$ 
are of the type $x_{\sigma}$ where $|\sigma|=i$;
\item {\em Completeness:} 
if  label $x_{\sigma}$ exists in $\TT$ then all labels $x_{\rho}$, $\rho\in 2^{\leq |\sigma|}$  exist in $\TT$;
\item {\em Uniqueness:} each string in $\TT$ can have at most one label;
\item {\em Consistency:} if  $\rho$ of level $\ell_k$ in $\TT$ has label $x_{\sigma}$ then
for each $i<k$, $\rho\restr_{\ell_i}$ has label $x_{\sigma\restr_i}$. 
\end{enumerate}
\end{defi}

The tree $\TT$ will be determined as the limit of a computable labelling process
$(\TT_s)$ which is canonical with respect to the given $(\ell_i)$, in the following sense.

\begin{defi}[Canonical labelling process]\label{HOAd1QkH}
A labelling process  $(\TT_s)$ is canonical with respect to an increasing
sequence $(\ell_i)$ if the following properties hold for all $s$.
\begin{enumerate}[\hspace{0.5cm}(1)]
\item {\em Structure:} the tree $\TT_s$ is structured with respect to $(\ell_i)$ ;
\item {\em Finiteness:} only strings of length at most $\ell_s$ can have a label in $\TT_s$;
\item {\em Persistence:} if  $\rho$ has label $x_{\sigma}$ in $\TT_s$, then it has the same label
in $\TT_t$ for all $t>s$.
\end{enumerate}
\end{defi}
Clearly a canonical labelling process $(\TT_s)$ has a limit, which is a structured 
partially labelled tree. From now on we suppress the qualification
`with respect to an increasing sequence $(\ell_i)$' when we use the notions of
Definitions \ref{oFUyKrVWUK} and \ref{HOAd1QkH}, and always assume the fixed sequence
$(\ell_i)$ that is given in Lemma \ref{Dedk3zkhlL}.

Note that Definition \ref{oFUyKrVWUK} and Definition \ref{HOAd1QkH} allow
the possibility that a single label $x_{\sigma}$ may have many copies at some level $\ell_k$ 
of some $\TT_s$.

\subsection{Definitions for the labelling construction}\label{HlOZPVjo9b}
The following notation will be useful.
\begin{defi}[Labelled subset and size]\label{i3ZIXrFRyy}
Given a  structured partially labelled tree $\TT$, let $\TTast$ denote the set which includes the empty string and all 
labelled strings in $\TT$. 
The length of the longest $\sigma$ such that a label $x_{\sigma}$ has been placed on a string in $\TTast$  is denoted $\dvls{\TTast}$. 
\end{defi}

The purpose of the labelling process is to ensure that
 for every string $\sigma$ there is eventually a string $\rho$ in $\TT$ which is extendible in $\PP$
and which has label $x_{\sigma}$. In this sense, the enumeration $(\QQ_s)$ of $\QQ$ is the main driver
of the process, and determines the placement of additional copies of already existing labels.
Timing is a crucial aspect of the labelling process, however, and for this reason
we will not use the arbitrary enumeration $(\QQ_s)$ directly in the construction. We use the following
filtered version instead, which takes into account the existing labelling at each stage.
 Here and in the following discussions, a leaf of $\TTast_s$ is a string in
$\TTast_s$ which does not have any labelled proper extensions.
 
 \begin{defi}[Filtered enumeration of $\QQ$]\label{PvKfvvp7j8}
During the construction we define 
a \ce set of strings $\DD$ inductively. $\DD_s$ denotes
the set of strings enumerated into $\DD$ by the end of stage $s$. 
\begin{itemize}
\item At stage 0 let $\DD_0=\emptyset$;
\item At stage $s+1$, if there exists a leaf of $\TTast_s$ 
which does not belong in $\DD_s$
and has a prefix in $\QQ_s$, pick the lexicographically least such leaf and enumerate it into $\DD$.
\end{itemize}
\end{defi}
Clearly $\dbra{\DD_s}\subseteq\dbra{\QQ_s}$ while the converse is not generally true.
Note that for a string $\rho$ to enter $\DD$ at stage $s+1$ it is not enough to have a prefix in $\QQ_{s}$.
Hence $(\DD_s)$ is a filtered version of $(\QQ_{s})$, 
in the sense that only {\em previously} labelled strings can be enumerated into $\DD_s$.

As remarked previously, Definition \ref{HOAd1QkH} crucially allows 
for the possibility that a single label $x_{\sigma}$ may have many copies at some level $\ell_n$ 
of some $\TT_s$. Amongst all of the strings with the same label $x_{\sigma}$, however, we shall ensure that at any given time there is precisely one of these strings which is given the special status of being \emph{active}. Roughly speaking, the active strings are those above which it presently seems there is still room for further coding at the next level. If $|\sigma|=n$ and the label $x_{\sigma}$ is placed on $\tau$, then while $\tau$ is active we may place labels for one element extensions of $\sigma$ on the extensions of $\tau$ of level $\ell_{n+1}$. We shall do so as the demands of the construction require, working from left to right. As each of these labels are placed, we do not have to be concerned initially as to whether they are placed on strings with prefixes in $\mathcal{Q}$ -- we simply place the labels and then wait for the enumeration of $\DD$ to subsequently alert us if we have placed labels on strings which do not have extensions in $\mathcal{P}$. Once labels have been placed on all extensions of $\tau$ of level $\ell_{n+1}$, $\tau$ is said to be \emph{saturated}. It should be noted that at any given point, if $\sigma \subset \sigma'$, $\tau$ and $\tau'$ have labels $x_{\sigma}$, $x_{\sigma'}$ respectively and are both active, it will not necessarily hold that $\tau \subset \tau'$. We make the following definitions.

\begin{defi}[Active strings]\label{bbsbTD6Vg6}
Given a canonical labelling process $(\TT_s)$, 
a string $\rho$ in $\TTast_s$ is {\em active} if it has some label $x_{\sigma}$ and $\rho$ was the last string
to receive this label in the approximations $\TT_0,\dots,\TT_s$.
\end{defi}
A string in $\TTast_s$ that is not active is called {\em inactive}. 
 
\begin{defi}[Saturated strings]
A string $\rho$ of level $\ell_k$ of a structured partially labelled tree $\TT$ is saturated if
all of its extensions at level $\ell_{k+1}$ of $\TT$ are labelled.
\end{defi}

Note that if $(\TT_s)$ is a canonical labelling process and a string of $\TT_s$ is saturated, 
then the same string will also be saturated in 
$\TT_{t}$ for all $t>s$. Similarly, by Definition \ref{bbsbTD6Vg6}, if a string  in $\TTast_s$ is inactive  then the same string will also be inactive in 
$\TTast_{t}$ for all $t>s$.

Each stage of the construction of $(\TT_s)$ after stage $0$, will be one of the following two kinds.
\begin{defi}[Expansionary and adaptive stages]
A stage $s+1$ is called {\em expansionary} if $\dvls{\TT^{\ast}_{s+1}}>\dvls{\TT^{\ast}_{s}}$. 
Otherwise $s+1$ is called an {\em adaptive} stage.
\end{defi}
It will be immediate from the construction that $s+1$ is expansionary if and only if $\DD_{s+1}=\DD_{s}$.

In the labelling construction we will explicitly deactivate strings 
in order to emphasize the newly inactive strings. It will be evident that
this is compatible with Definition \ref{bbsbTD6Vg6}.

\begin{defi}[Cloning a branch]\label{g2LjPVcPo}
Given $\delta,\beta\in\TTast_s$ such that $\delta$ is a leaf, suppose that:
\begin{enumerate}[\hspace{0.5cm}(i)]
\item if $x_{\sigma}, x_{\tau}$ are the labels of $\beta,\delta$ respectively then 
$\sigma\subset\tau$;
\item $\eta$ is the leftmost string of length $|\delta|$ which extends $\beta$
and $\eta\restr_{\ell_{k+1}}$ is not labelled.
\end{enumerate}
{\em Cloning $\delta$ above $\beta$} means to
label $\eta\restr_{\ell_i}$ with
the label of $\delta\restr_{\ell_i}$,
for each $i$ such that $\ell_i\in (|\beta|,|\delta|]$, making each of these strings active.
\end{defi}

In Definition \ref{g2LjPVcPo}, we allow the case that $\beta$ is the empty string $\lambda$, in which case there is no label placed on $\beta$.
Given a labelled string $\rho$ in $\TT_s$, the {\em active clone} of $\rho$ in $\TT_s$
is the unique active string  in $\TT_s$ which has the same label as $\rho$.
Note that  the active clone of an active string is the string itself.
For uniformity, we define the active clone of the empty string $\lambda$ to be $\lambda$. 

\subsection{The labelling construction}\label{CHnceHLowy}
At stage 0 we place a label $x_{\lambda}$ on the leftmost string of length $\ell_0$ and make this string active.
 At stage $s+1$ suppose that the labelled tree $\TT_s$ has been defined, and consider the
following two cases:

{\em Expansionary stage:}
If $\DD_{s+1}=\DD_s$ then let $\TT_{s+1}\restr_{\ell_{s}}=\TT_s\restr_{\ell_{s}}$
and for each active leaf $\rho$  of $\TT^{\ast}_{s}$ with label some $x_{\sigma}$, place
labels $x_{\sigma\ast 0},x_{\sigma\ast 1}$ on the leftmost and rightmost extensions of $\rho$ of level 
$\ell_{s+1}$, making these strings active,  then end stage $s+1$.

{\em Adaptive stage:}
If $\DD_{s+1}\neq\DD_s$ then let $\delta$ be the  string in $\DD_{s+1}-\DD_s$ and let 
$\alpha_j, j\leq k$ be the empty or labelled initial segments of $\delta$ in order of magnitude, so that $\alpha_0=\lambda$ and $\alpha_k=\delta$.  
Also let $\beta_j, j\leq k$ be the
active clones of $\alpha_j, j\leq k$ respectively in $\TT_s$. 
Let $j_0$ be the largest number $j<k$ such that $\beta_j$ is not saturated and 
\begin{itemize}
\item deactivate $\beta_{j}$ for each $j\in (j_0, k]$;
\item clone $\delta$ above $\beta_{j_0}$.
\end{itemize}
If such $j_0$ does not exist, 
say that the  construction {\em terminates} at stage $s+1$; 
otherwise end stage $s+1$.

\section{Properties of the labelling algorithm}\label{EgBPR7blWw} 
Note that since $(\ell_i)$ is increasing, each string of length $\ell_k$ 
has at least two distinct extensions of length $\ell_{k+1}$. Hence
the expansionary stages of the construction are well-defined. A straightforward induction on stages  
suffices to establish that  $(\TT_s)$ is a canonical labelling process, according to
Definition \ref{HOAd1QkH}. In particular, the placing of labels satisfies the consistency condition required in order to define a valid functional. While Definition \ref{bbsbTD6Vg6} specifies the active strings at each stage, during the construction we have also directly deactivated strings, as well as activating them 
during the process of cloning and at expansionary stages. It is clear that at any stage the strings which have been activated and not directly deactivated by the construction, are precisely those which are active according to Definition \ref{bbsbTD6Vg6}, since it is precisely when we place a new version of a given label that we deactivate the previously active string with that label.  It also follows by a straightforward induction on stages, that at the end of each stage $s$,  any leaf of $\TTast_s$ is either active, or else has already been enumerated into $\DD_s$. In particular, when $\delta$ is enumerated into $\DD_{s+1}$ during stage $s+1$,  it was previously active and is deactivated during this  adaptive stage. This means, in the notation of Section \ref{CHnceHLowy}, that when we deactivate $\beta_k$ without requiring that it be saturated, in fact  $\beta_k=\delta$, so that the only strings which are deactivated during an adaptive stage are strings which are enumerated into $\DD$, or else are saturated: 
\begin{equation} \label{jwyefg} 
\parbox{12cm}{Inactive labelled strings
in $\TT_s$ are either saturated or else belong to $\DD_s$.}
\end{equation}
The following is also established easily by induction on stages: 
\begin{equation}\label{oYIMPg9qd}
\parbox{12cm}{If $\delta\in\DD$ then for all $s$, no proper extension of $\delta$ is
labelled in $\TT_s$.}
\end{equation}

\subsection{Non-termination}
In order to show that the labelling construction does not terminate (i.e. that we do not run out of room for coding),  it suffices to establish that $\lambda$ is never saturated (regarding $\lambda$ as of level $\ell_{-1}$ in the definition of saturation). The following definition will be useful. 

\begin{defi}[Set of active strings]
Let $U_s$ be the set of active strings in $\TT_s$. For each string $\rho$ 
let $U_s(\rho)$ be the set of strings $\gamma\supseteq\rho$ which are active in $\TT_s$.
\end{defi}

We are interested in the weight of the active strings, where the weight of a set of strings $V$ is defined by:
\[
\wgt{V}=\sum_{\eta\in V} 2^{-|\eta|}.
\]
In order to show that $\lambda$ is never saturated we shall first establish: 
\begin{equation}\label{pbW1St2Z5Q}
\parbox{8cm}{$\wgt{U_s}+\mu(\DD_s)< 1$ for all stages $s$.}
\end{equation}
The following claim will also be established by induction on stages: 
\begin{equation}\label{sTEHuzW1Ac}
\parbox{13cm}{Given any $s$ and any infinite binary sequence $Z$ which does not have a prefix in $\DD_s$,
the largest labelled initial segment of $Z$ is active in $\TT_s$.}
\end{equation}
Note that in this statement it is possible that $Z$ does not have a labelled initial segment,
in which case the assertion is trivially true.
An immediate consequence of \eqref{sTEHuzW1Ac} is that
\begin{equation}\label{6k1gwTROBL}
\parbox{13cm}{For each $s$ and each $\nu$ which is labelled in $\TT_s$, 
we have $\dbra{\nu}\subseteq \dbra{\DD_s}\cup \dbra{U_s(\nu)}$.}
\end{equation}
Now if  $\lambda$ is saturated at stage $s$ then 
the entire Cantor space is covered by the labelled strings of length $\ell_0$.
Hence by \eqref{6k1gwTROBL}
we have
$2^{\omega}\subseteq \dbra{\DD_s}\cup \dbra{U_s}$.
Then 
$1\leq \mu(\DD_s)+\wgt{U_s}$, which contradicts \eqref{pbW1St2Z5Q}.

It remains to establish (\ref{pbW1St2Z5Q}) and (\ref{sTEHuzW1Ac}).   To see (\ref{pbW1St2Z5Q}), note first that at each stage $s$ and for each $\sigma$ 
there is at most one active string in $\TT_s$ with label $x_{\sigma}$.
Since for each $n$ there are only $2^n$ strings of length $n$, we have:
\[
\wgt{U_s}\leq \sum_{\rho\in U_s} 2^{-|\rho|}\leq
\sum_n \left(\sum_{\eta\in U_s\cap 2^{\ell_n}} 2^{-|\eta|}\right)\leq
\sum_n \left(2^n\cdot 2^{-\ell_n}\right)=\sum_n 2^{n-\ell_n}.
\]
If we combine this with our hypothesis \eqref{bi1VeHzCL9} 
we get $\wgt{U_s}< \mu(\PP)=1-\mu(\QQ_s)$.
By the fact
$\dbra{\DD_s}\subseteq\dbra{\QQ_s}$ which we observed after Definition \ref{PvKfvvp7j8}, we get
$\wgt{U_s}< 1-\mu(\DD_s)$, from which (\ref{pbW1St2Z5Q}) follows.

It remains to prove \eqref{sTEHuzW1Ac} by induction on the stages of the labelling construction.
At stage $0$ we have $\DD_0=\emptyset$ and all labelled strings are active.
It follows that in this case \eqref{sTEHuzW1Ac} holds.
Inductively suppose that \eqref{sTEHuzW1Ac} holds at stage $s$.
If stage $s+1$ is expansionary, then no string is deactivated, and any new labels
are placed on strings that become active. So given an infinite binary sequence $Z$ and the largest
initial segment $\nu$ of $Z$ that is labelled in $\TT_{s+1}$, either the label of $\nu$
existed in $\TT_{s}$ or it did not.
In the first case we can use the inductive hypothesis to conclude that $\nu$ is active in  $\TT_{s+1}$.
In the second case we can conclude the same, due to the fact that newly labelled strings in expansionary stages
are active.
Hence if $s+1$ is expansionary,
\eqref{sTEHuzW1Ac} continues to holds at stage $s+1$.

Now suppose that  $s+1$ is an adaptive stage and let $\delta$ be the unique
element of $\DD_{s+1}-\DD_s$. Also let $Z$ be an infinite binary sequence which has at least one labelled
initial segment in $\TT_s$, and let $\nu$ be the largest such initial segment. If $\nu=\delta$ there is nothing to prove, so assume
otherwise. If no initial segment of $Z$ is deactivated during stage $s+1$, the claim follows by the induction hypothesis.
For the remaining case, let $\eta$ be the largest labelled prefix of $Z$ which is deactivated during stage $s+1$.
If $|\eta|=\ell_k$, let $\eta'=Z\restr_{\ell_{k+1}}$. Since $\eta$ was deactivated at $s+1$, it follows from (\ref{jwyefg}) that it was saturated
in $\TT_s$. This means that $\eta'$ must be labelled in $\TT_s$. Hence the largest labelled initial segment of $Z$
in $\TT_s$, which is also the largest in $\TT_{s+1}$, is active in 
$\TT_{s+1}$, just as it was active in $\TT_s$.
Hence \eqref{sTEHuzW1Ac} holds for $Z$ at stage $s+1$.
Finally consider the case where $Z$ did not have a labelled initial segment in $\TT_s$, but it does in $\TT_{s+1}$.
Since all the newly labelled strings at stage $s+1$ are active in $\TT_{s+1}$, in this case also we can conclude that
\eqref{sTEHuzW1Ac} holds for $Z$ at stage $s+1$.

This completes the induction step and the proof of 
\eqref{sTEHuzW1Ac}.

\subsection{Growth of the tree and the enumeration of \texorpdfstring{$\DD$}{D}}
Now that we have proved the construction does not terminate, the rest of the verification is essentially routine. Note that at each stage $s$ the set $\TTast_s$ is finite, 
and that in every adaptive stage some previously unlabelled strings receive labels. Since $(\TT_s)$ is
a canonical labelling process, it follows from the fact that the  labelling construction does not terminate that there are infinitely many expansionary stages. Hence:
\begin{equation}\label{G87wYdPFY}
\lim_s \dvls{\TT_s}=\infty.
\end{equation}
Recall that $\TTast$ is the limit of all $\TTast_s$. We wish to show that:
\begin{equation}\label{aPXewefgugw9}
\parbox{11cm}{If $\tau\in \TTast$ extends a string in $ \QQ$ then there exists $s$ with $\dbra{\tau} \subseteq  \dbra{\DD_s}$.}
\end{equation}

 This will follow once we establish the following fact: 
 
 \begin{equation}\label{aPXewfkrugw9}
\parbox{12.5cm}{ If $\tau\in \TTast_s$ is active, 
there is at least one leaf of $\TTast_s$ extending $\tau$ which is not in $\DD_s$. }
\end{equation}

In order to see that (\ref{aPXewefgugw9}) follows from (\ref{aPXewfkrugw9}), suppose that $s_0$ is the least stage at which $\tau\in \TTast_{s_0}$ and there exists $\tau'\subseteq \tau$ with $\tau'\in \QQ_{s_0}$. Let $s_1$ be the least expansionary stage $>s_0$. At the beginning of stage $s_1$, (\ref{aPXewfkrugw9}) implies that no string $\tau''\supseteq \tau$ in $\TTast_{s_1}$ can be active, meaning that all such strings must either be saturated or else belong to $\DD_{s_1}$, by (\ref{jwyefg}). This implies that $\dbra{\tau} \subseteq \dbra{\DD_{s_1}}$ as required. 

We establish (\ref{aPXewfkrugw9}) by induction on stages. If $s_0$ is the first stage at which $\tau$ is active, then no extensions of $\tau$ are in $\DD_{s_0}$. At any subsequent stage $s>s_0$ at which $\tau$ is still active, if a leaf $\delta$ extending $\tau$ is enumerated into $\DD_s$, then that leaf will be cloned above some $\beta_{j_0}\supseteq \tau$, which completes the induction step.

\section{The coding process and its verification}\label{yPyARDIifv}
In this section we show how to determine the code $Y$ of a given infinite binary sequence $X$,
so that $X\restr_n$ can be uniformly computed by $Y\restr_{\ell_n}$. We define this reduction based on the labelling process of Section \ref{mtzrcubXzt} and its properties.
Note that a direct consequence of the labelling construction
of Section \ref{CHnceHLowy} is that every leaf of $\TTast_s$ has the same length. 


\subsection{The coding process}\label{UvsqmTLQ3}

For any given $X$ consider the downwards closure of (i.e.\ the set of all initial segments of)  strings that are labelled with a prefix of $X$. Since there are infinitely many expansionary stages, this set is an infinite tree, and so has an infinite path by K\"{o}nig's Lemma. We let $Y$, the code for $X$, be any such infinite path.   

\subsection{The coding verification}
We verify that the code $Y$ of $X$ determined by the above construction has the required properties.

\paragraph{$Y$ belongs to $\PP$}{}
It suffices to show that $Y$ does not have a prefix in $\QQ$.
This follows by  \eqref{aPXewefgugw9} and  (\ref{oYIMPg9qd}). 

\paragraph{$Y$ computes $X$ with oracle use $(\ell_n)$}{
We show that for each $n$ we can compute
$X\restr_n$ uniformly from $Y\restr_{\ell_n}$. Given $n$ and $Y\restr_{\ell_n}$ we simply
run the labelling construction until the first stage $s_0$ where the string $Y\restr_{\ell_n}$ is labelled 
in $\TT_{s_0}$. 
Note that since $Y\in [\TTast]$ and $\TT$ is a structured 
partially labelled tree, such a stage exists. 
If $x_{\sigma}$ is the label of $Y\restr_{\ell_n}$ then
$X\restr_{n}=\sigma$.}

This concludes the verification of the coding process and
the proof of Lemma \ref{Dedk3zkhlL}.

\end{document}